\documentclass[twocolumn,prd,showpacs,preprintnumbers,amsmath,amssymb,floatfix,aps]{revtex4}
\usepackage{float}
\usepackage{amsmath}
\usepackage{amsfonts}
\usepackage{amssymb}
\usepackage[dvips]{graphicx}
\usepackage[usenames]{color}

\usepackage{bbm}
\def\beq{\begin{equation}}
\def\eeq{\end{equation}}
\def\bea{\begin{eqnarray}}
\def\eea{\end{eqnarray}}
\def\ba{\begin{array}}
\def\ea{\end{array}}

\def\part{\partial}

\begin{document}

\preprint{UdeM-GPP-TH-07-161}
\preprint{arXiv:0710.3236 [hep-lat]}

\title{Phase transitions in a 3 dimensional lattice loop gas}
\author{Richard MacKenzie$^a$}
%\email{rbmack@lps.umontreal.ca}
\author{F. Nebia-Rahal$^a$}
%\email{faizanr@lps.umontreal.ca}
\author{M. B. Paranjape$^{a,b}$}
%\email{paranj@lps.umontreal.ca}

\affiliation{$^a$Groupe de physique des particules, D\'epartement de
physique, Universit\'e de Montr\'eal, C.P. 6128, Succ. Centre-ville,
Montr\'eal, Qu\'ebec, Canada, H3C 3J7 }
\affiliation{$^b$Center for Quantum Spacetime, Department of
Physics, Sogang University, Shinsu-dong \#1, Mapo-gu,
Seoul, 121-742, Korea}

\begin{abstract}
We investigate, via Monte Carlo simulations, the phase structure of a system of closed, non-intersecting but otherwise non-interacting, loops in 3 Euclidean dimensions.  The loops correspond to closed trajectories of massive particles and we find a phase transition as a function of their mass.  We identify the order parameter as the average length of the loops at equilibrium.  This order parameter exhibits a sharp increase as the mass is decreased through a critical value, the behaviour seems to be a cross-over transition.  We believe that the model represents  an effective description of the broken-symmetry sector of the 2+1 dimensional abelian Higgs model, in the extreme strong coupling  limit.   The massive gauge bosons and the neutral scalars are decoupled, and the relevant low-lying excitations correspond to vortices and anti-vortices.  The functional integral can be approximated by a sum over simple, closed vortex loop configurations.  We present a novel fashion to generate non-intersecting closed loops, starting from a tetrahedral tessellation of three space.   The two phases that we find admit the following interpretation: the usual Higgs phase and a novel phase which is heralded by the appearance of effectively infinitely long loops.  We compute the expectation value of the Wilson loop operator and that of the Polyakov loop operator.  The Wilson loop exhibits perimeter law behaviour in both phases implying that the transition corresponds neither to the restoration of symmetry nor to confinement.  The effective interaction between external charges is screened in both phases, however there is a dramatic increase in the polarization cloud in the novel phase as shown by the energy shift introduced by the Wilson loop.    
\end{abstract}

\maketitle

\section{Introduction} Although it is one of the simplest gauge theories, the abelian Higgs model is of substantial theoretical interest \cite{Anderson,Rajantie}.   It corresponds to scalar electrodynamics consisting of a charged scalar field and a neutral vector field which is the gauge boson of a $U(1)$ local gauge symmetry.   It can be defined in any number of space-time dimensions.  

In 1+1 dimensions,  the general absence of spontaneous symmetry breaking \cite{HCMW} poses a puzzle as to the manifestation of the $U(1)$ symmetry for the naively spontaneously  broken sector.  Indeed, topological vortices play the role of instantons and give rise to tunnelling transitions which end up disordering the vacuum \cite{DHN}.  The symmetry is actually restored; however, a $U(1)$ gauge theory in 1+1 dimensions is classically linearly confining.  Consequently, charged states are hidden in neutral bound states.  

In 2+1 dimensions, the compact version of the theory behaves quite differently than the non-compact version.  A $U(1)$ gauge theory can be thought of as a theory either with gauge group $U(1)$ living on the compact manifold $S^1$, or with gauge group $R^1$, (the real numbers under addition) living on the non-compact manifold $R^1$.  The compact theory, in the unbroken phase, shows linear confinement of charges, instead of the classically expected logarithmic potential \cite{Polyakov}, due to magnetic monopoles which act as instantons.   The actual details of the mechanism of this confinement are rather complicated and we will not describe them here.  In the non-compact case, magnetic monopoles do not exist; hence the expression of the symmetry should be along more traditional lines: either the symmetry is manifest with a logarithmic potential between charged particles, or it is spontaneously broken and the interaction is screened.  In principle, the theory could even be linearly confining for fractionally charged external sources.  

The classically spontaneously broken sector of the non-compact theory in 2+1 dimensions will be of interest in this article.  Here,  the theory has topological solitons, Nielsen-Olesen \cite{Nielsen} vortices, which tend to disorder the vacuum. Vortex lines in a type II superconductor \cite{Rajantie} are examples of  physical phenomena which are well described by such solitons.  Looking at the 3-dimensional Euclidean version of the theory,  these vortices extend to tubes of quantized magnetic flux.  For these configurations to have relevance to the functional integral, finiteness of the action requires that they form closed loops.    The contribution of such closed vortex loops to the expectation value of the Wilson loop\cite{Wilson} was computed, at strong coupling, in a heuristic semi-classical analysis by Samuel \cite{Samuel}.  There it was proposed that, if the vortices are light enough, they should effectively condense, giving rise to a novel phase, what was called the ``spaghetti vacuum.''  What this means is that contributions to the Euclidean functional integral come preponderantly from configurations which are full of vortex loops.  It was further deduced that there should be a logarithmic potential  induced between external charges.  Such a potential is in fact confining: it takes an infinite amount of energy to move two particles infinitely far away from each other, although it is not {\em linearly} confining.   A phase transition going from the standard symmetry broken phase to a novel phase corresponding to a disgorging of vortex tubes into the vacuum has also been proposed by Einhorn and Savit \cite{einhornsavit} in their study of the lattice abelian Higgs model.  

In this paper we study, by means of Monte Carlo simulations on a lattice \cite{Wilson,Creutz2}, a discretized, {\em effective} version of the abelian Higgs model.  This amounts to the study of a gas of loops on a lattice with Boltzmann weight corresponding to the total length of the loops.  We find indeed that there is a rather sharp transition from small average loop length to a configuration with an effectively infinite loop, the average loop length showing a remarkable  increase.  This kind of transition is very reminiscent of percolation type transitions\cite{percolation}. In the Abelian Higgs model interpretation the transition is from the standard Higgs phase to a novel phase characterized by saturation of the functional integral by configurations that are filled with vortex flux loops.  We do not however find the corresponding classical logarithmic potential induced between external charges \cite{Samuel}.  External charges are still screened; however, we find that the energy of the screening cloud increases dramatically.

\section{Effective abelian Higgs model at strong coupling}
The abelian Higgs model is described by the Lagrangian density
 \begin{equation}
 \mathcal{L}=-\frac{1}{4} F_{\mu\nu}F^{\mu\nu}+\frac{1}{2}D_{\mu}\phi (D^{\mu}\phi)^{*} -\frac{\lambda}{4} (|\phi|^{2}-\eta^{2})^{2},
\end{equation}
 where $\phi$ is a complex scalar field, $A_{\mu}$ is a $U(1)$ gauge field, $D_{\mu}=\partial_{\mu}-ieA_{\mu}$, $ F_{\mu \nu}=\partial_{\mu} A_{\nu} -\partial_{\nu} A_{\mu}$ $(\mu,\nu=0,1,2) $ and $\lambda,e $ and $\eta$ are taken to be positive constants.  This theory undergoes spontaneous symmetry breaking appended by the Higgs mechanism yielding a perturbative spectrum of a massive vector boson with mass $M=e\eta$ and a neutral scalar boson with mass $m=\sqrt{2\lambda} ~\eta$.  
 
Additionally, the theory contains vortex solitons of quantized magnetic flux in this sector.  Their mass behaves like $\mu =\eta^2\times f(2\lambda/e^2)$ \cite{jr}, where $f(2\lambda/e^2)$ is a function that satisfies $f(1)=1$, but can take any positive value as a function of $\lambda/e^2$.    We can take the strong coupling limit $\lambda, e\rightarrow\infty$ while keeping $\eta$ and $\lambda/e^2$ fixed.  This decouples the perturbative excitations, $m,M\to\infty$,  leaving only the vortices as the effective excitations.  As was shown in \cite{Nielsen}, in this limit, the size of the vortices vanishes and their world lines resemble perfect, fundamental strings.  We will only study the abelian Higgs model in the description afforded by this effective model.   The phase structure of the effective model must be the same as that of the original abelian Higgs model sufficiently deep in the strong coupling regime.  Thus our results will shed light on the asymptotic region of the strong coupling limit of the abelian Higgs model.  

In the lowest approximation, neglecting gradient energies due to curvature, the action is given by $\mu\times L$ for a closed loop of length $L$.  The Euclidean vacuum-to-vacuum amplitude is obtained by functionally integrating over field configurations that correspond to the following Euclidean time histories:  they are the classical vacuum configuration at the initial time,  contain  a number of virtual vortex anti-vortex pairs at intermediate steps, and revert back to the classical vacuum configuration at the final Euclidean time. Thus, in this limit,  the abelian Higgs model is equivalent to a gas of massive particles that carry a conserved flux; these particles are non-interacting except when they are in close proximity.  All other excitations and their interactions are negligible.  Thus the functional integral is evaluated by integrating over field configurations corresponding to closed vortex loops \cite{Samuel}, indeed a 3 dimensional loop gas.  We will  calculate  this integral by a numerical Monte Carlo simulation on a lattice discretized version of this effective theory.   

\section{Lattice loops} On the lattice, it is not straightforward how to construct closed loops.  We construct closed loops by starting with a tessellation of Euclidean 3-space with (non-regular) tetrahedra.  To generate this tessellation, we start with a  body-centered cubic (bcc) lattice in a box of size $ N = N_{s}^{2}N_{\tau} $.  
%$N_{\tau}\sim k_B\beta$ the inverse temperature.  
Joining the central vertex in each cube with its corners  fills each cube with 6 identical pyramids with square bases given by the faces of each cube.  Splitting each pyramid in half yields two irregular tetrahedra and the desired tessellation.   To define the splitting, we start with the cube with one vertex at the origin, extending into the positive octant.  We cut each face from the origin to the opposite diagonal corner in the $(x,y)$  $(y,z)$ and the $(z,x)$ plane respectively.  Then we translate this scheme throughout the lattice.  This converts each pyramid into two (identical) non regular tetrahedra, giving a total of 12 tetrahedra in each cube.  All points inside the box fall into one tetrahedron or another, except for the set of measure zero which resides on the surfaces of the tetrahedra. Therefore we have filled space with tetrahedra.  

Loops are generated by distributing the three cube roots of unity over the vertices of the tessellation.  A given triangular face is associated with an oriented length of vortex tube piercing it and going to the center if the change of phase about the triangle corresponds to $\pm 2\pi$, using the right hand rule.  If a triangular face of a given tetrahedron has the cube roots of unity distributed on the vertices so that one unit of flux enters the tetrahedron, with a little reflection it is easy to see that assigning any cube root of unity to the fourth vertex of the tetrahedron necessarily defines one unit of flux exiting the tetrahedron through another of its triangular faces, passing from the center of the original tetrahedron to the center of the neighbouring one.  But since space is filled with tetrahedra this exiting flux tube enters the neighbouring tetrahedron, and repeating the argument it must exit this tetrahedron, entering a third tetrahedron, and so on. It is topologically impossible for the vortex line to end; it must ultimately close on itself, forming a closed loop, since the lattice is made up of only a finite number of tetrahedra.  The resulting configuration is a system of closed vortex  loops which are by construction non-intersecting, since it is also topologically impossible for two vortices to enter the same tetrahedron.  A similar scheme  was originally implemented on a cubic lattice \cite{Vilenkin} for cosmic strings.  There, one could have two vortex lines entering a cube and two exiting it, although it was impossible to resolve the path of the vortex lines inside the cube.  Our tetrahedral dissection of the cube resolves this ambiguity.  
 
The actual Euclidean geometrical length of the loop will depend on the explicit trajectory that the loop takes since the distances between the centers of neighbouring tetrahedra are not all the same.   For a long loop, these geometrical factors average out, simply giving a renormalization of the value of $\mu$, which includes the lattice spacing.    The corresponding effective action is $\mathrm{S}_{\rm eff} \,  \propto \,  \left( \mu \times L_{T} \right)$ where the $L_T$ is the number of triangles through which the loop passes, where by abuse of notation we use the same symbol $\mu$ to represent the mass times the lattice spacing.  We will call $L_T$ the length of the loop.  The shortest closed loop has $L_T=4$ while the maximum is $ L_{T, \rm max} = 12 N_{s}^{2}N_{\tau}$.

\section{Monte Carlo simulations} Our simulations are performed on a bcc cubic lattice with $ N_{s}=100$, $ N_{\tau}=100 $ and $ \mu$ from 0 to 1.5 using Monte Carlo simulations \cite{Binney,Landau}.  We begin with an initial arbitrary configuration of closed loops.  Then we use the standard Metropolis algorithm to generate an ensemble of configurations which follow the  Boltzmann distribution with weight given by $ \left( e^{-\mu L_{T}} \right)$. 

\subsection{Thermalization}
\begin{figure}
%\centerline{\includegraphics[bb= 10 17 269 227,width=0.9\linewidth]{LT0.pdf}}  
\centerline{\includegraphics[width=0.9\linewidth]{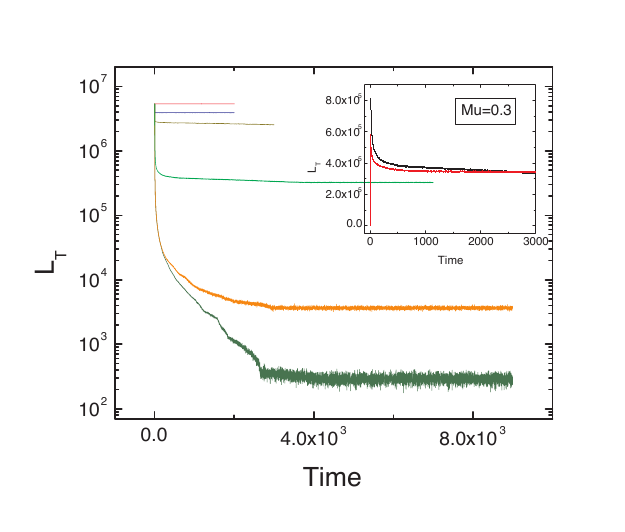}} \caption{\label{fig:L0} (color online) Total length of loops as a function of time for $ 100^{3}$ bcc lattice. From top to bottom, $\mu=0,0.1,0.15,0.3,0.9,1.5$.  Insert displays $\mu =0.3$, for two different initial configurations.}
\end{figure}

 In Fig.~\ref{fig:L0}, on a semi logarithmic scale, we show the convergence of the total length of loops $L_{T}$ versus Monte Carlo time (updates) for several values of $\mu$ and $ N_{\tau}=100$. Our unit of time corresponds to one complete update of each site of the $100^{3}$ bcc lattice. The equilibrium state does not depend on the initial state, but it is strongly dependent on $\mu$.  The average total length  and the absolute fluctuations grow  as $\mu$ is decreased, but the relative fluctuations diminish.

\subsection{Numerical evidence for a change in phase} 
\begin{figure}
 %\centerline{\includegraphics[bb= 15 20 266 223,width=0.9\linewidth]{lt2.pdf}}
 \centerline{\includegraphics[width=0.9\linewidth]{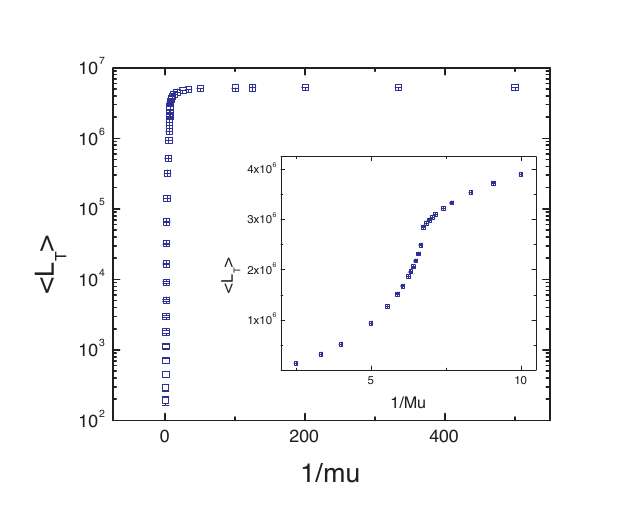}}
 \caption{\label{fig:Lmoy} (color online) Expectation value of the total length of loops as a function of $ \mu$, $0\leq\mu \leq 1.5$, with the insert focussing on the region of the transition.}
\end{figure}

In Fig.~\ref{fig:Lmoy}, we show the expectation value of the total length of loops $\left\langle L_{T}\right\rangle$ as a function of inverse $\mu$ on a logarithmic scale.
We see that  there is a dramatic change in the curve around $\mu=0.15$ indicating a transition in the system.
We define the total density of the loops $\rho$ as the ratio of the computed $L_T$ to $ L_{T, \rm max}$.  The transition corresponds to the appearance of effectively infinite loops in the simulation.  If the simulation could be done in infinite space, at the transition a truly infinite loop would appear.  Infinitely long loops\cite{Vilenkin} in a finite volume are operationally defined as those loops having a length $L$, much longer than that they would normally have if they corresponded to a closed, self-avoiding random walk.  The size of a self-avoiding random walk on a simple cubic lattice behaves as $ L^{\sim 3/5}$, closing the loop adds one constraint which should not greatly modify this scaling law.   Actually it is found that the exponent should be slightly less than $3/5$ on a cubic lattice, but the exact value is not analytically known\cite{ms}.  On our $100^3$ bcc lattice, with the tetrahedral trajectories, it is not clear how the size of self avoiding, closed  random walks would exactly scale with their length.   However, taking the cubic lattice result as a guide,  simple calculation yields $200^{5/3}\approx  6840$, where 200 is the number of steps from one side to the other of our lattice, hopping from the corner of a cube to the center, back to the next corner and so on.  Therefore we treat any loop of length substantially greater than 10000 as an infinite loop.   

\begin{figure}
\centerline{\includegraphics[bb= 239 281 492 570,width=0.60\linewidth]{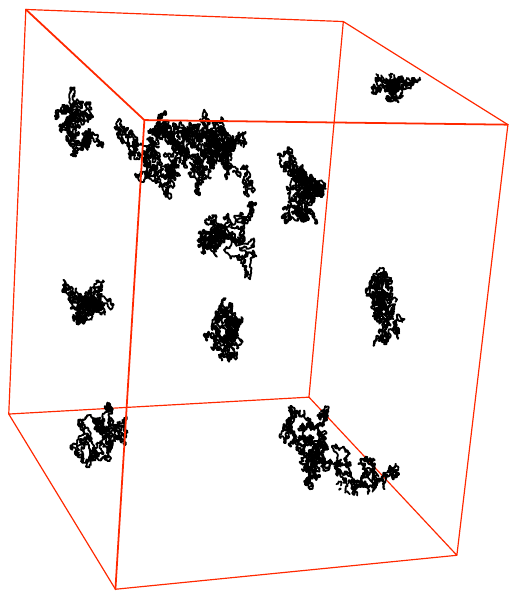}}
%\centerline{\includegraphics[width=0.90\linewidth]{Oct08_6.pdf}}
%\centerline{\includegraphics[bb= 8 18 278 214,width=0.9\linewidth]{densite.pdf}}
\centerline{\includegraphics[bb= 239 281 492 580,width=0.60\linewidth]{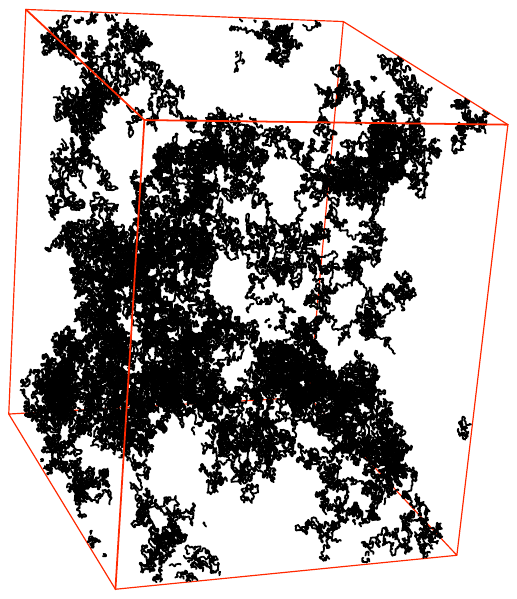}}
%\centerline{\includegraphics[width=0.60\linewidth]{Oct08_7.pdf}}
\caption{(color online) Snapshots of the closed loops generated in equilibrium phase for $100^{3}$ bcc lattice. Identification of the axes is arbitrary. (Top): Only the largest closed loops are displayed for $\mu=0.152$, $1400\leq L\leq3000$. (Bottom): Larger loops are formed after the transition (see text) as shown for $\mu=0.148$; the closed loop shown has $L\simeq155500$.}
\label{fig:densiteInf}
\end{figure}

 Fig.~\ref{fig:densiteInf} shows snapshots of closed vortex loops, with periodic boundary conditions,  generated in the equilibrium phase before the transition, for $\mu=0.152$ (top) where only finite closed loops are present, and after the transition, for $\mu=0.148$ (bottom) where larger closed loops are formed. For a better visualization, only some loops are presented, as otherwise the picture looks completely black, filled with vortex loops.
 There is some theoretical understanding of this phenomenon in thermodynamical studies of a network of cosmic strings. The authors of \cite{mt} have noted that at formation, the density of states of cosmic strings in the early Universe is dominated by a large loop containing most of the energy with a thermal distribution of finite, low mass,  strings.  This density distribution was also described by the authors in \cite{Vilenkin}.  The number of microstates available to the system is much greater when reorganized as a large number  of finite  loops augmented by one infinite loop.   The density at which this happens corresponds to the Hagedorn temperature and the transition corresponds to the Hagedorn phase transition \cite{h}.    In the cosmological situation, as the Universe expands the phase space favours configurations where all the strings are chopped off into the smallest possible loops. 

\begin{figure}
%\centering
% \centerline{\includegraphics[bb= 8 18 278 214,width=0.9\linewidth]{densite.pdf}}
 \centerline{\includegraphics[width=0.9\linewidth]{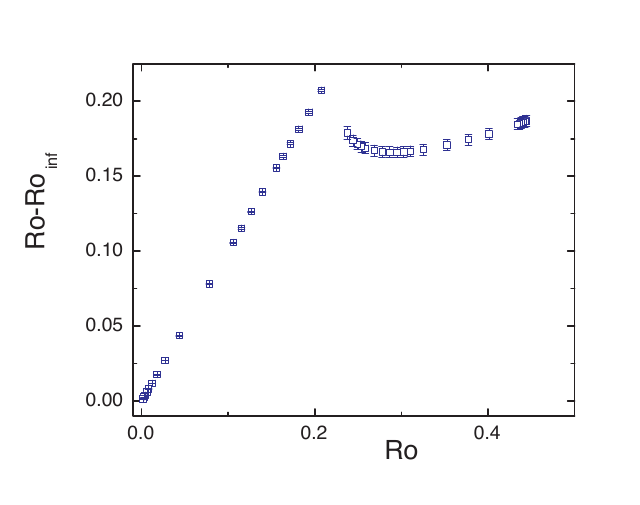}}
 \caption{\label{fig:densite} (color online) Density of finite loops as a function of the total density for $0\leq\mu \leq 1.5$.}
\end{figure}

In Fig.~\ref{fig:densite}, we graph the density of finite loops.   For small values of the total density, there are no infinite loops; hence the curve is linear with slope 1.  We see a dramatic transition around $ \rho=0.207$ which corresponds to $\mu =0.152$.  At the transition there is a sudden reorganization of the vortex loops into one infinitely long loop and a number of finite loops.    Remarkably, the increase in the total density/length of loops caused by further decreasing $\mu$ occurs only by appending to the infinitely long loop, the density of finite vortex loops remaining essentially constant.
\section{Order parameters} 
We want to analyze the nature of the novel phase and to study the system around the transition point.  For that, we turn to the following observables as order parameters:  the average length of the loops, the Wilson loop operator \cite{Wilson} and the Polyakov \cite{Polyakov} loop operator.  
\subsection{Length of the average loop}
The average length of the loops, $\langle L\rangle$, that make up the ensemble of equilibrium configurations shows a remarkable transition as a function of $\mu$.  Below we plot $\langle L\rangle$ as a function of $\ln\mu$.  
 \begin{figure}[h]
 \centerline{\includegraphics[width=1.1\linewidth]{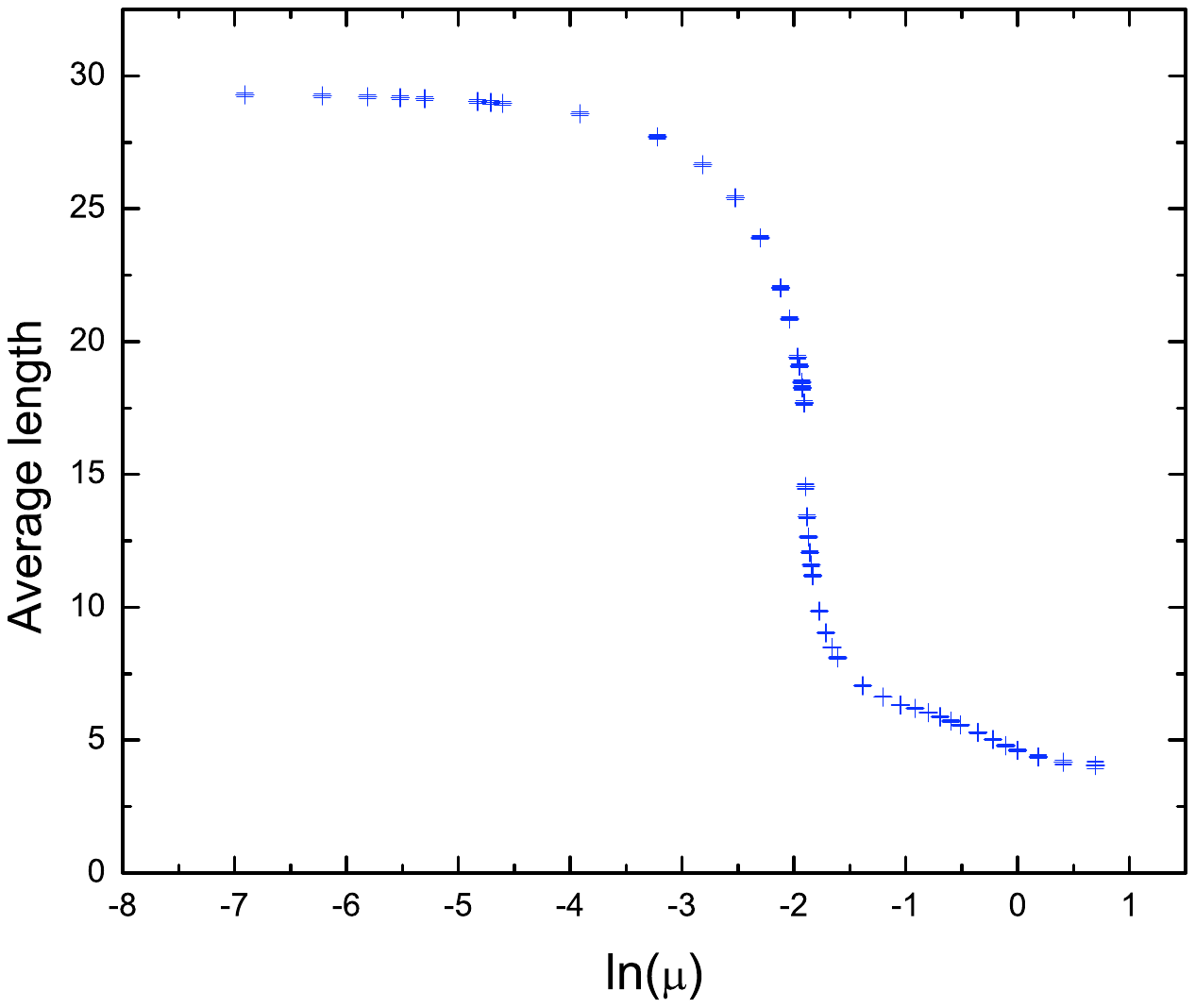}}
  \caption{\label{LM_lnMu}(color online) Average length of the loops as a function of $\ln(\mu)$.}
\end{figure}
We observe that the transition occurs at $\ln(\mu)\approx -1.9\approx\ln(0.15)$, exactly at the point where the effectively infinite loop appears.  Continuing the figure to larger values of $\ln(\mu)$ is not feasible as the equilibrium configuration is not easily obtained.  The Monte Carlo process of attaining equilibrium is asymptotically slowed.

\subsection{Wilson loop}
The Wilson loop operator corresponds to inserting into the system two static, equal but opposite charges  $q$, separating them by a distance $L$ for a duration $T$ with $(T\gg L)$, and then annihilating them. The expectation value of Wilson loop operator is given by
 \beq
 W \left( L,T \right)=\left< e^{-i\frac{q}{e} \oint A_{\mu} dx_{\mu} } \right>,
 \label{wilson}
\eeq
where the integration is over the rectangular Wilson loop $(L\times T)$ contour.  For our effective model a dramatic simplification occurs, $\oint A_{\mu} dx_{\mu}$ exactly measures $2\pi$ times the linking number $\nu$ of the Wilson loop with the closed vortex loops:  
\beq
 W \left( L,T \right)=\left\langle e^{ -i(2\pi q/e)\nu } \right\rangle.
\eeq
 For large $T$, 
$
W \left( L,T \right) \sim e^{-\Delta\left( L\right)\cdot T},
$
where $\Delta\left( L\right)=\lim_{T\rightarrow\infty}-(1/T)\ln (W(L,T))$, the energy shift,  is the interaction energy of the static $q\bar{q}$ pair separated by a distance $L$ \cite{Wilson}.  In the usual Higgs phase, we expect that finite closed vortex loops will give a perimeter behaviour for the expectation value of Wilson loop operator, which means that the charges are screened.  In the novel phase, however, the infinitely long vortex loops could give a contribution that has no relation to the perimeter.
 \begin{figure}[h]
 \centerline{\includegraphics[width=0.9\linewidth]{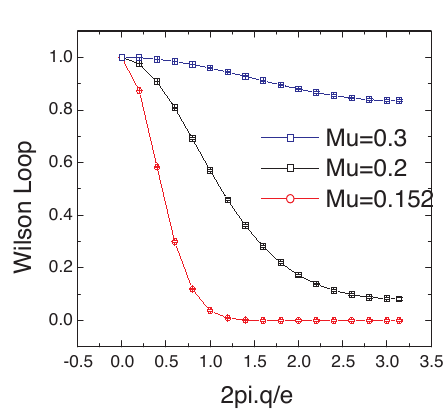}}
   %\end{figure}
 %\begin{figure}[h]
 \centerline{\includegraphics[width=0.9\linewidth]{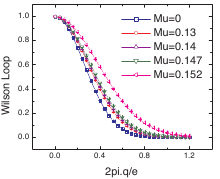}}
  \caption{\label{5au}(color online) Wilson loop $W(L=20,T=80)$ as a function of $2\pi q/e$ from 0 to $\pi$ for $\mu=0.3,0.2,0.152$ (upper)and $\mu=0,0.13,0.14,0.147, 0.152$. (lower)}
\end{figure}
 \begin{figure}[h]
 \centerline{\includegraphics[width=0.9\linewidth]{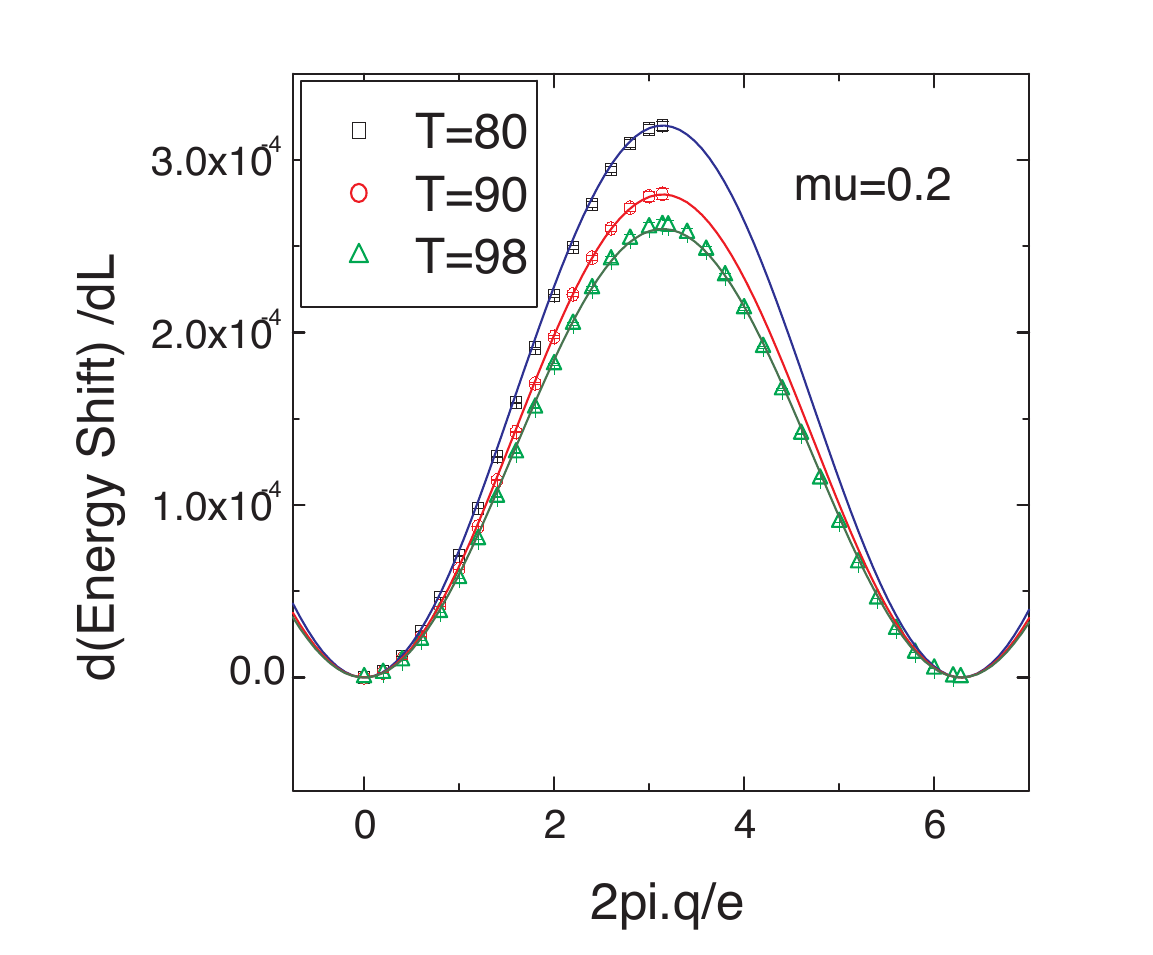}}
  \caption{\label{5b}(color online)  $d\Delta/dL$ as a function of $2\pi q/e$ for $T=80,90,98$.  The data points are our numerical results for  $\mu=0.2$. The solid lines are the fit of the form $C(T)\sin^{2}((2\pi q/e)/2)$.}
  \end{figure}
 \begin{figure}[h]
  \centerline{\includegraphics[width=0.9\linewidth]{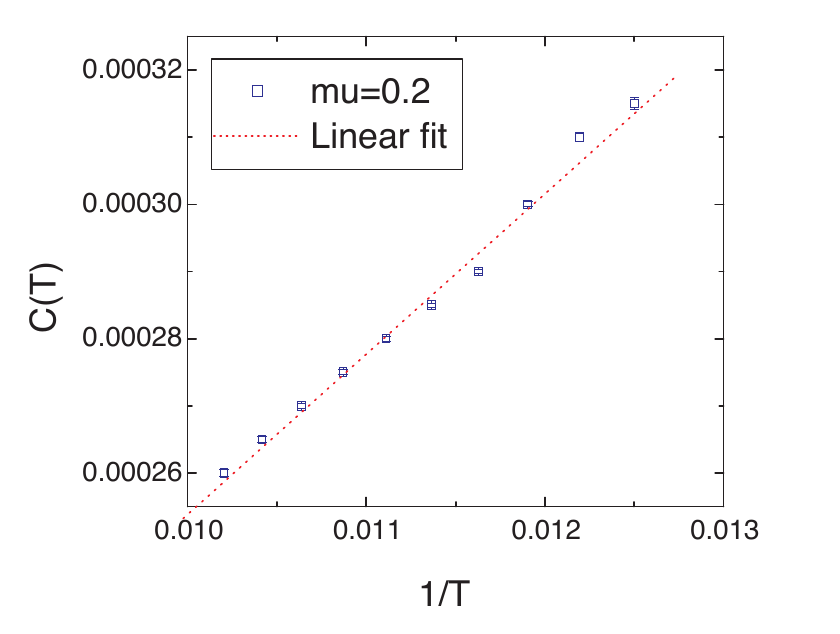}}
 \caption{\label{5c}(color online) $C(T)$ as a function of $1/T,\,\,\, \mu=.2$. The dotted line is the linear fit}
\end{figure}

Figures~\ref{5au} - \ref{5c} shows our results for the numerical calculation of the potential between $q\bar q$.    In Fig.~\ref{5au}  we start with the calculation of the Wilson loop operator $W(L,T)$  for $0\leq 2\pi q/e \leq 2\pi$ for various values of $\mu$ in the Higgs phase in the upper graph for $\mu=.3$ to the border of the phase transiition at $\mu=.152$ and in the lower graph in the novel phase for $\mu=.152$ to $\mu=0$.  We note that the curve moves rapidly from the borderline value at $\mu = 0.152$ to that deep in the Higgs phase at $\mu = 0.3$, conversely there is very little movement from the transition at $\mu=.152$ to $\mu=0$.

The energy shift at finite $T$, $\Delta(L,T)$, is analyzed for the  value $\mu =.2$ in the Higgs phase, shown in the Fig.~\ref{5b}.  The energy shift should be a periodic function of the external charge $q$; the expected form is $\sim \sin^{2}((2\pi q/e)/2)$  \cite{Coleman}.  A perimeter law then would imply 
\beq
\Delta(L,T) = A(T)\cdot\sin^{2}((2\pi q/e)/2)(L+T)/T,
\eeq
while an area law would give 
\beq
\Delta(L,T) = B(T)\cdot\sin^{2}((2\pi q/e)/2) (L\times T)/T.
\eeq
In general, we allow a sum of the two behaviours and   $A(T)\to A$ and $B(T)\to B $ independent of $T$ as $T\to\infty$.  Then 
\bea
d\Delta(L,T)/dL\to (A/T+B)\sin^{2}((2\pi q/e)/2)\\
\equiv C(T)\sin^{2}((2\pi q/e)/2)
\eea
should be independent of $L$.  The dots correspond to our numerical simulation; the solid lines correspond to the fit.  In Fig.~\ref{5c}, the $T$ dependence of  $C(T)$ is displayed; it is a linear function of $1/ T$.  The extrapolation of the curve to $T\to\infty$ yields $C(\infty)=0$, {\it i.e.} $B=0$.  This means that there is only the perimeter law behavior for the Wilson loop operator in the Higgs phase.

\begin{figure}[h]
 \centerline{\includegraphics[width=\linewidth]{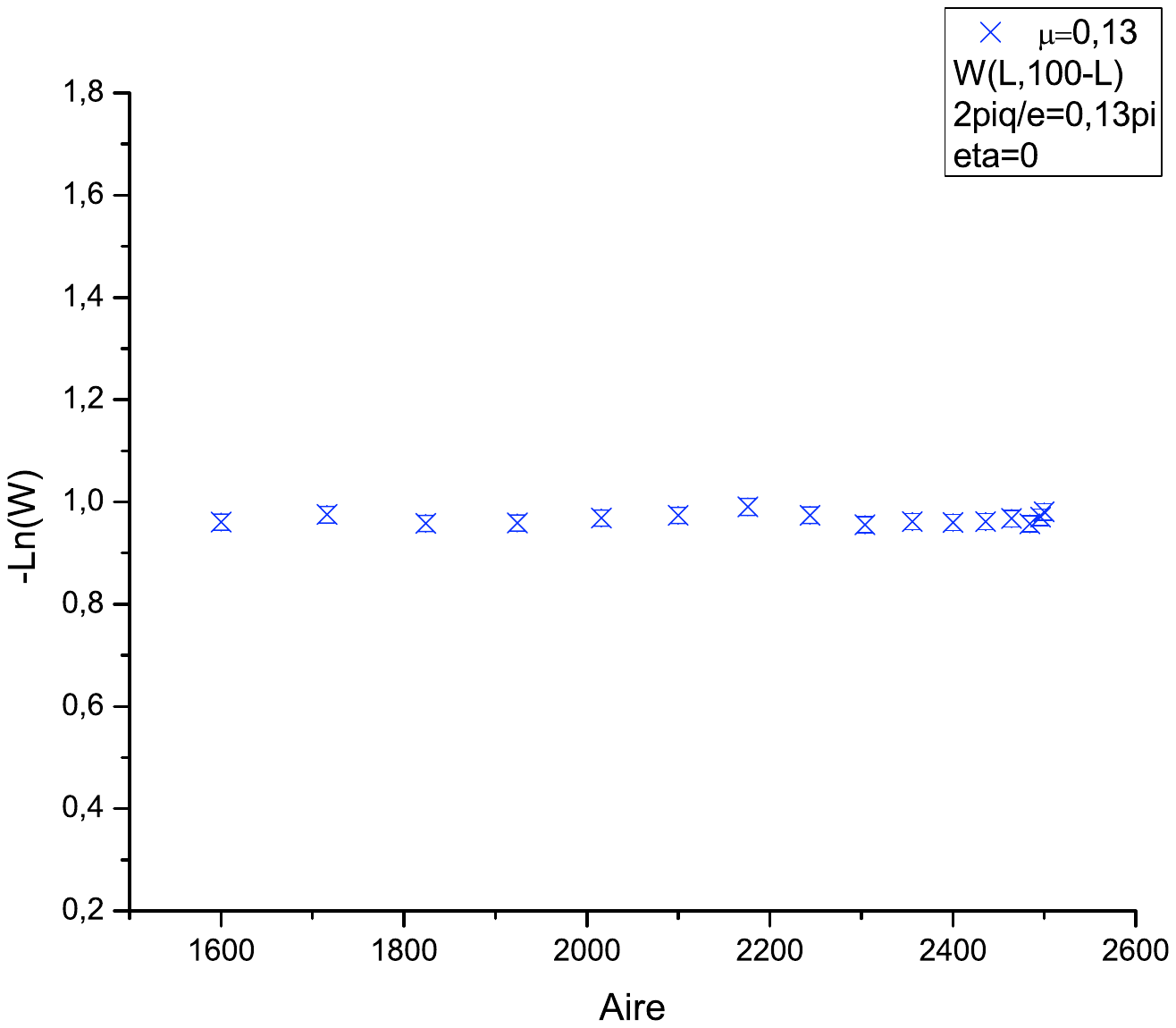}}
 \caption{\label{aire} (color online) Wilson loop $W(L,T=100-L)$ as a function of area for $\mu=.13$.}
\end{figure}

In the novel phase, for $\mu \leq 0.152$, the Wilson loop, as a function of $2\pi q/e$, does not vary greatly with $\mu$.   It decreases as $x\equiv 2\pi q/e$ approaches approximately $0.98$ where it vanishes.  This implies that the interaction energy of the external charges is so large that our numerical analysis is not able to resolve its value, within the resolution permitted by our lattice approximation.  It does not by any means imply confinement.  In the novel phase we cannot use the simple function $\sin^{2}((2\pi q/e)/2)$ to give the dependence on $q$.  However we can easily see that the Wilson loop is independent of the area.  In Fig.~\ref{aire} we plot the Wilson loop as a function of $L$ for a loop of size $(L,100-L)$ ie. for a fixed value of the perimeter, and for a fixed value $2\pi q/e=.4$.  It is evident that the value of the Wilson loop does not vary. 

\begin{figure}[h]
\includegraphics[width=0.95\linewidth]{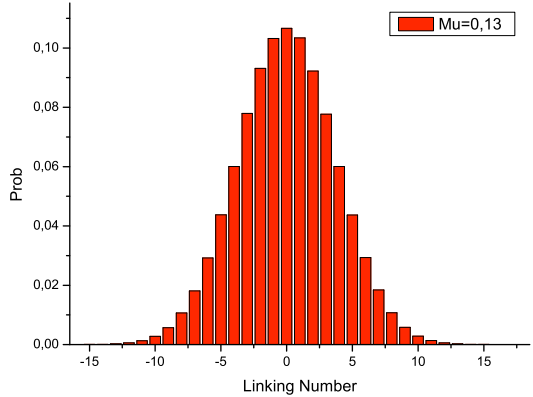}
 \caption{\label{h13} (color online) Histogram of the Wilson loop linking number $\nu$ for $W(L,T=40,80)$  for $\mu=.13$ in the novel phase.}
\end{figure}

\begin{figure}[h]
\includegraphics[width=0.95\linewidth]{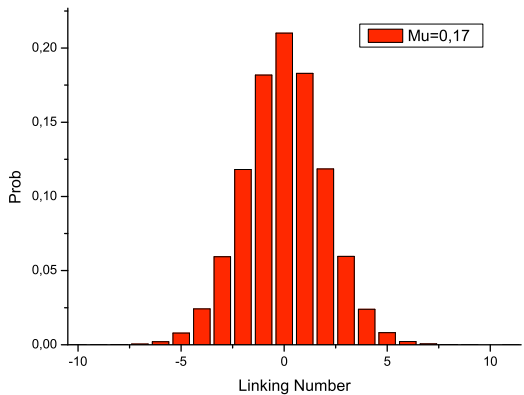}
 \caption{\label{h17} (color online) Histogram of the Wilson loop linking number for $W(L,T=40,80)$  for $\mu=.17$ in the Higgs phase.}
\end{figure}

\begin{figure}[h]
\includegraphics[width=0.95\linewidth]{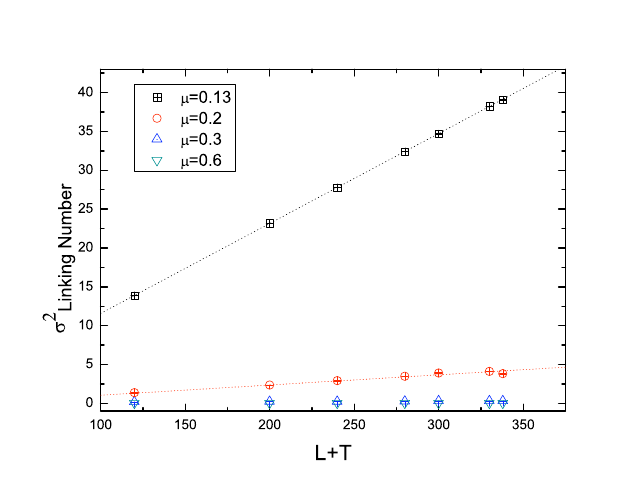}
 \caption{\label{sigma2} (color online) Plots of the variance as a function of the perimeter for $\mu=.13, .2,.3$ and $.6$.  The dashed lines are a linear fit.}
\end{figure}

To compare the novel phase with the Higgs phase, we can look at the distribution of the linking number, $\nu$, of the Wilson loop.  The histogram for the distribution of the linking number are given in Figures \ref{h13} and \ref{h17} for $\mu=.13$ and $\mu=.17$ respectively. 
Such histograms were calculated for different values of $\mu$ on either side of the transition and for different sizes of the Wilson loop.  In Fig. \ref{sigma2} we plot the variance $\sigma^2$, of the histograms as a function of the perimeter, $L+T$ for $L=40$ and $T$ varying.  Clearly we find a perimeter law for the variance.

The expectation value of Wilson loop is constructed from the histograms of linking number of the Wilson loop.  We compute $e^{i(2\pi q/e)\times \nu}$ for each value of the linking number $\nu$, weigh that phase with the number of configurations with that value of $\nu$, and then sum over all linking numbers.  This actually corresponds to calculating the characteristic function of the probability distribution function for the linking number  \cite{cf}.  Clearly this gives a sum of phases which are distributed over the unit circle in a manner depending on the exponent.  If the $\sigma\times 2\pi q/e$ is of the order of $o (1)$, the phases are essentially randomly distributed over the unit circle, and the characteristic function vanishes.  This is seen in Fig. \ref{5au} (lower graph) where the expectation value of the Wilson loop crashes to zero for $2\pi q/e\sim .6$ in the novel phase, when the variance suddenly becomes large.  This behaviour is to be contrasted with that in the Higgs phase where the expectation value is a smooth sinusoidal function of $2\pi q/e\in[0,2\pi ]$ as seen in Fig. \ref{5au} (upper) for $\mu>.152$.

\begin{figure}[h]
\includegraphics[width=0.95\linewidth]{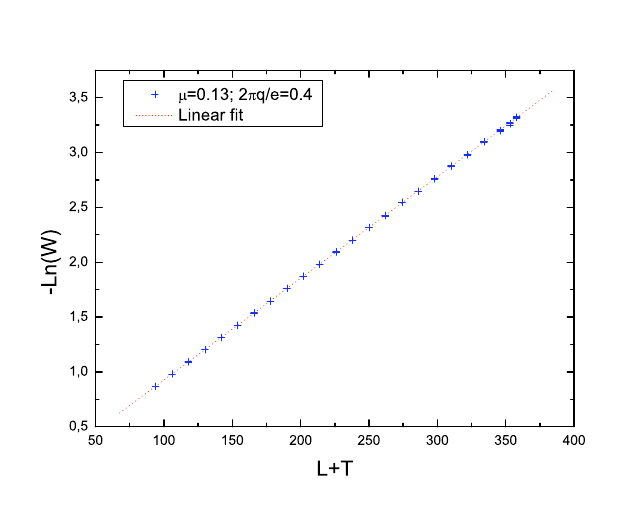}
 \caption{\label{pl1} (color online) Plot of $-\ln (W)$ for $\mu=.13$.  The dashed lines are a linear fit, we find $-\ln (W)=.009280(\pm4\times 10^{-6})\cdot(L+T)+.00350 (\pm 5\times10^{-4})$}
\end{figure}

\begin{figure}[h]
\includegraphics[width=0.95\linewidth]{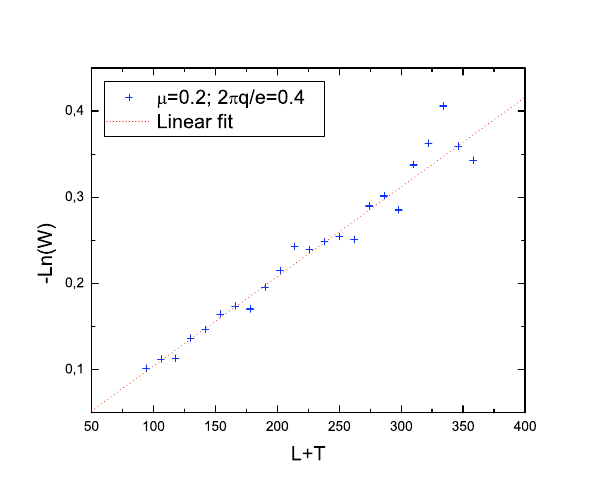}
 \caption{\label{pl2} (color online) Plot of $-\ln (W)$ for $\mu=.2$.  The dashed lines are a linear fit, we find $ -\ln (W ) =   0,001030        \pm  (3 \times 10^{-7})(L + T )+8,7\times 10^{-4}\pm (5\times 10^{-5})$}
\end{figure}

The perimeter law for the variance does translate into a perimeter law for the Wilson loop.  In Fig. \ref{pl1} and \ref{pl2}, for $\mu=.2$ in the Higgs phase and for $\mu=.13$ in the novel phase,  we plot the the log of the expectation of the Wilson loop as a function of the perimeter.  We see a perimeter law for the Wilson loop and a linear behaviour of the derivative of the energy shift in both phases.  We do our analysis for a fixed value of $2\pi q/e =.4$.


\begin{figure}[h]
\includegraphics[width=0.95\linewidth]{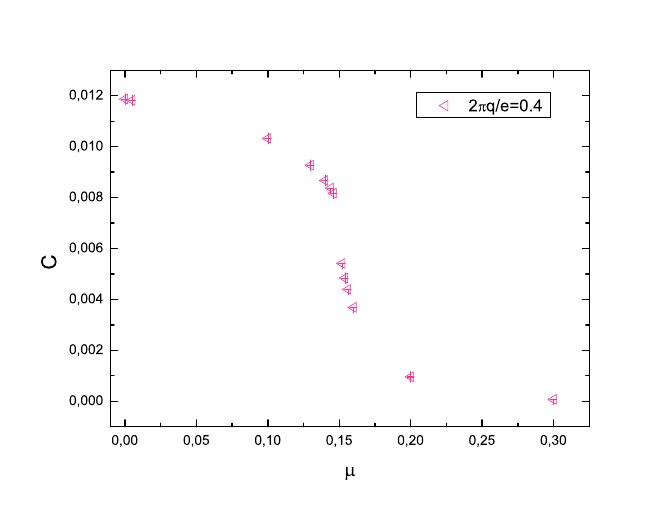}
 \caption{\label{Pente_L+T} (color online) Plot of the parameter $C$ as a function of $\mu$ at fixed $q$.}
\end{figure}

The conclusion we can make is that there is a dramatic increase in the polarization cloud surrounding the external charges as one passes from the Higgs phase to the novel phase.  From Fig.~\ref{pl1} and \ref{pl2} we see that the coefficient for the perimeter law for $\mu=.13$ in the novel phase is approximately 9 times larger than that for $\mu=.2$.  Indeed, we can construct graphs analogous to Figs. \ref{pl1}, \ref{pl2} for many values of $\mu$ and  plot the parameter $C$, which is the slope of the line in the graph of $-\ln (W)$ versus $L+T$.  We find a sharp cross over at the transition as shown in Fig. \ref{Pente_L+T}, indicating the almost ten-fold increase of the polarization cloud energy.  

\subsection{Polyakov loop}
\begin{figure}[h]
 \centerline{\includegraphics[width=0.95\linewidth]{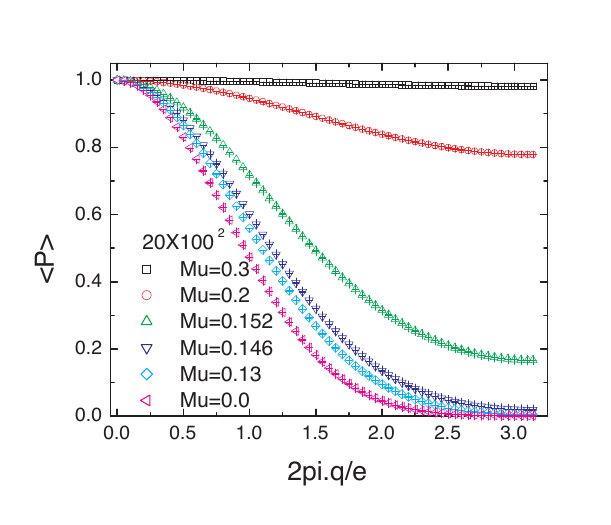}}
 \caption{\label{polyakov} (color online) Polyakov loop for various values of $\mu$ for a lattice 20$\times 100^2$.}
\end{figure}

At finite temperatures, one looks at the behaviour of the Polyakov loop operator, which is defined as the Wilson loop variable taken along the entire (periodic) time direction $N_{\tau}$ for a fixed spatial position $\vec{x} $.  This is related to the free energy of the system, $F_{q}$, in the presence of a single heavy quark by \cite{Rothe}:
$ \left\langle P \left( \vec{x} \right) \right\rangle = e^{-\beta F_{q}}.$    In Fig.~\ref{polyakov}, we see the behavior of the expectation value of the Polyakov loop operator $\left\langle P \right\rangle$ mirrors almost exactly the behaviour of the Wilson loop as in Fig.~\ref{5au}.   
The position of the transition has changed to a smaller value of $\mu$, and at a larger value of $2\pi q/e$as should be expected,  because the temperature has been increased, corresponding to a Euclidean time direction of length 20.

\section{Scaling study} 

\begin{figure}[h]
\centerline{\includegraphics[width=\linewidth]{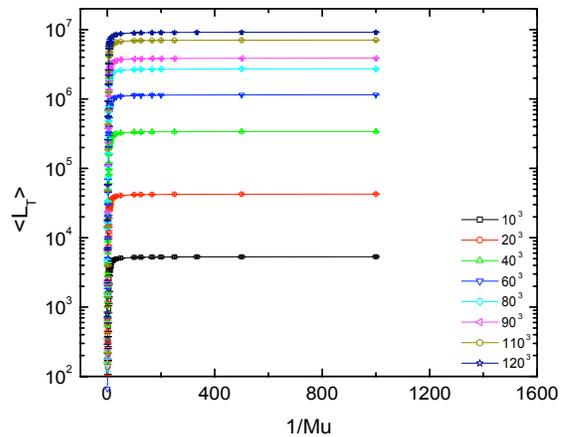}}
\caption{\label{sc1} (color online) Mean total length of loops $\langle L_T\rangle$ for various lattice sizes.}
\end{figure}

\begin{figure}[h]
\centerline{\includegraphics[width=\linewidth]{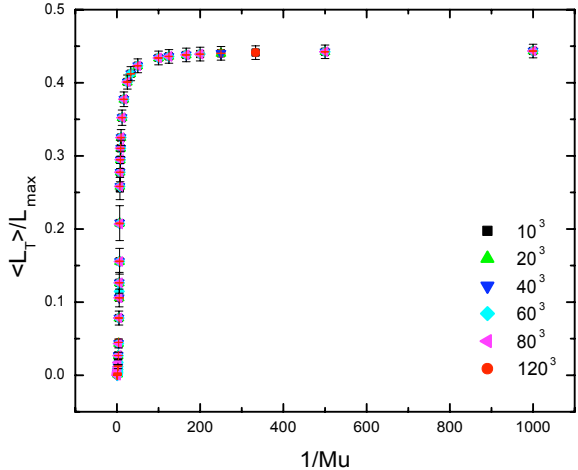}}
\caption{\label{sc2} (color online) Normalized  mean total length of loops $\langle L_T\rangle/L_{max}$ for various lattice sizes.}
\end{figure}

In any lattice simulation, it is important to use a lattice which is sufficiently large to eliminate finite size effects. In Fig.~\ref{sc1}, the mean total length of loops as a function of $\mu$ is displayed for various lattice sizes; normalizing by the maximum possible length of loops Fig. \ref{sc2}, we see that it is independent of the lattice size, as expected. We find that the transition point occurs for  $\mu\approx 0.15$ for all lattices larger than $10^3$, so the lattice size used in this study ($100^3$) is amply sufficient.

\begin{figure}[h]
\centerline{\includegraphics[width=\linewidth]{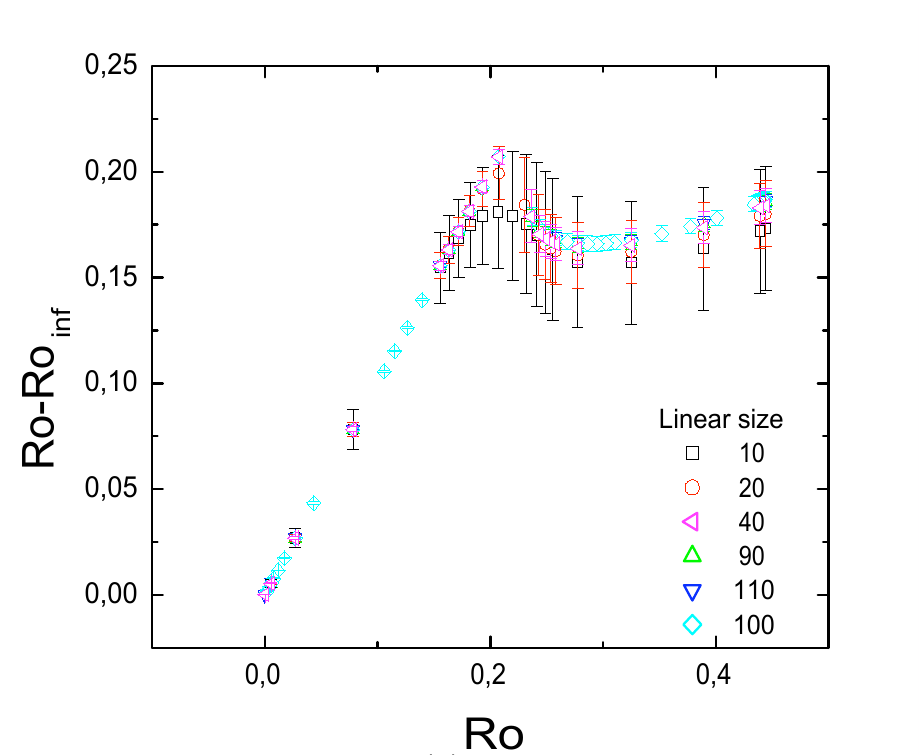}}
\caption{\label{sc3} (color online) Density of finite loops as a function of the total density for various lattice sizes.}
\end{figure}

\begin{figure}[h]
\centerline{\includegraphics[width=\linewidth]{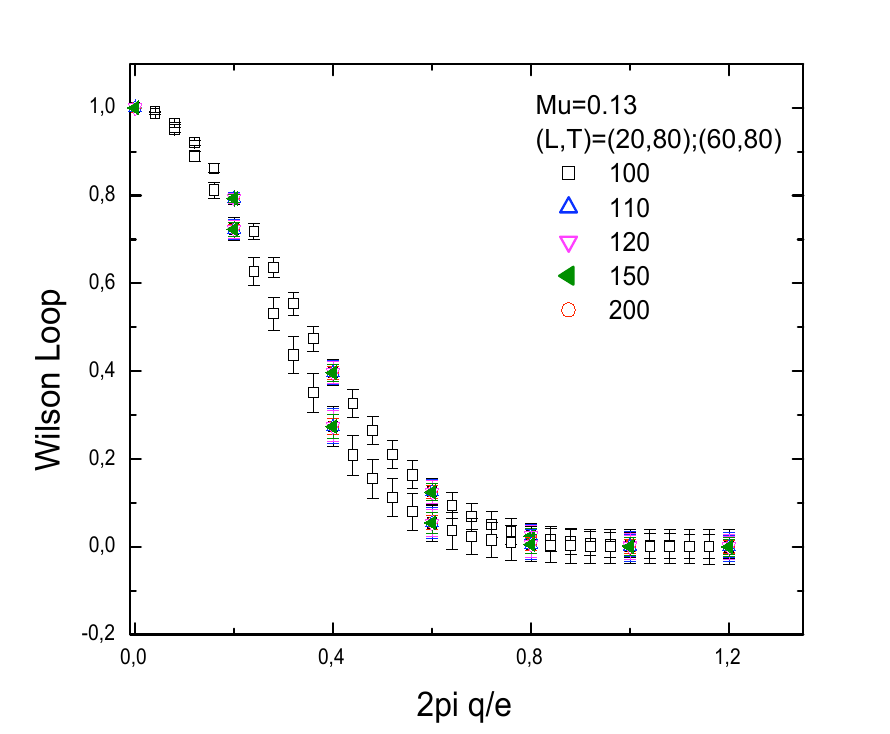}}
\caption{\label{sc4} (color online) $W(L,T)$ for $\mu=0.13$ for various lattice sizes.}
\end{figure}

In Fig.~\ref{sc3}, the density of finite loops versus the total density is illustrated, for various lattice sizes.  The large error bars are only present for the smaller lattices sizes, $10^3, 20^3$, already at $40^3$ we approach a consistent size independent density as a function of $\mu$.  For lattices sizes greater than $\approx 90^3$, the results are essentially independent of the lattice size.  
Finally in Fig.~\ref{sc4} the expectation value of two sizes of Wilson loop are displayed for various lattices sizes for $\mu=0.13$ as a function fo $2\pi q/e$.  Again, the results are clearly independent of the lattice size.  
Therefore we conclude that using a lattice considerably larger than $40^3$, and especially the size $100^3$ that was used for most of the simulations are perfectly adequate to remove all finite size effects.  

\section{Discussion and conclusions} Our results  show numerical evidence for a novel phase in the phase diagram of the 3-d abelian Higgs model at the asymptotic boundary corresponding to strong (infinite) coupling in the spontaneously broken Higgs phase.  At strong coupling the perturbative massive excitations, corresponding to the gauge boson and the neutral scalar, are completely decoupled.  The only remaining particle is the vortex, which is an adjustable parameter.  For a large mass of the vortex, the vacuum configuration is saturated by short loops of virtual vortex anti-vortex pairs. As this mass is lowered, the vacuum is filled with longer and longer virtual vortex anti-vortex pair loops.  Finally at a critical value, there is a transition to a novel phase, in which the vortex loops reorganize into one effectively infinite loop in addition to a bath of smaller finite loops.    

In the Higgs phase, external charges are screened due to a polarization cloud which leads to a  perimeter law for the Wilson loop, external charges are not confined.   Smaller than a critical value for the mass of the vortices, the polarization cloud increases dramatically causing the energy shift as defined by the Wilson loop to increase 9-10 fold.  The Wilson loop behaviour however, remains the perimeter law, contrary to the behaviour that was surmised in \cite{Samuel}.  Since we have decoupled all perturbative excitations of the scalar field, including specifically charged excitations, it is in principle possible for the Wilson loop to exhibit linear confinement.  We explicitly find no dependence on the area for the Wilson loop.  We find that the Wilson and Polyakov loop order parameters both vanish after a large enough value of the external charge, however this simply means that our lattice calculation is not able to resolve the details of its behaviour.  

The novel phase is characterized by the appearance of an effectively infinite vortex loop.  The individual finite vortex loops suddenly reorganize at the transition into one infinite loop and a gas of remaining finite loops, as a function of $\mu$.  The total length of the loops increases as a function of decreasing vortex mass, primarily through appending to the infinite vortex loop.  This novel phase was predicted by \cite{Samuel} and also in \cite{einhornsavit}.  In \cite{einhornsavit} the transition is described as the passage between the phases labelled VII and I.  

In \cite{fradkinshenker} it was proven that there is no transition between the Higgs phase and the Coulomb phase, however, they look strictly at the compact version of the model.  We are considering the non-compact model here, the analysis for non-existence of confinement does not apply.  Unfortunately, we do not find any evidence of confinement, even for fractionally charged external charges.  Numerical work similar to ours was also undertaken by \cite{cgp}, however they did not see our phase.  There are two differences between our simulation and theirs. First, we simulate only the effective model; hence, our results are valid only on the asymptotic boundary of the full phase diagram of the Abelian Higgs model whereas they \cite{cgp}, simulate directly the gauge fields and scalar fields.  Second, the limits taken are slightly different: we take $\lambda ,e\to\infty$ keeping $\lambda/e^2$ fixed, whereas they take first $\lambda\to\infty$ and only afterwards do they take $e\to\infty$.  Hence we go to the corner of the phase diagram (in $\lambda$ and $e$ space) along a particular fixed line, while they go up to the rectangular edge first (at $\lambda=\infty$) and then move along the edge (taking $e\to\infty$) to the corner.  It is clearly possible the two limiting procedures do not commute. Furthermore, their numerical work was done on a $16^3$ lattice, which, according to our scaling analysis, most probably will exhibit finite size effects.   The existence of the novel phase is expected to have important ramifications for the phase structure of the model in the presence of the Chern-Simons term \cite{ip}.  

\section{Acknowledgments} We thank NSERC of Canada and  the Center for Quantum Spacetime of Sogang University with grant number R11-2005-021   for financial support.  We thank Yvan Saint-Aubin, Jeong-Hyuck Park, and David Ridout for useful discussions and we thank Jacques Richer, of the R\'eseau qu\'ebecois de calcul de haute performance (RQCHP), Montr\'eal,  for paramount help with the programming and its implementation.  We also thank the  RQCHP for allocating computer time.  Finally we thank the Perimeter Institute for Theoretical Physics and the (Kavli) Institute of Theoretical Physics of the Chinese Academy of Sciences for hospitality while the paper was being revised.

%%%%%%%%%%%%%%%%%%%%  BIBLIO  %%%%%%%%%%%%%%%%%%%%%%%%%%%

\end{document}